\documentclass[prl,twocolumn,superscriptaddress,showpacs,preprintnumbers,amsmath,amssymb]{revtex4}
\usepackage{amssymb}

\usepackage{graphicx}
\usepackage{dcolumn}
\usepackage{bm}
\usepackage{epsfig}
\usepackage{amsmath}
\newcommand{\ignore}[1]{}

\begin{document}


\title{Giant Rashba-Spin Splitting of Bi(111) Bilayer on Large Band Gap
$\beta-$In$_2$Se$_3$}

\author{Wenmei Ming}
\affiliation{Department of Materials Science and Engineering,
University of Utah, Salt Lake City, UT 84112}

\author{Z. F. Wang}
\affiliation{Department of Materials Science and Engineering,
University of Utah, Salt Lake City, UT 84112}

\author{Feng Liu}
\thanks{Corresponding author. E-mail: fliu@eng.utah.edu}
\affiliation{Department of Materials Science and Engineering,
University of Utah, Salt Lake City, UT 84112}

\date{\today}


\begin{abstract}
Experimentally it is still challenging to epitaxially grow Bi(111)
bilayer (BL) on conventional semiconductor substrate. Here, we
propose a substrate of $\beta-$In$_2$Se$_3$(0001) with van der Waals
like cleavage and large band gap of 1.2~eV. We have investigated the
electronic structure of BL on one quintuple-layer (QL)
$\beta-$In$_2$Se$_3$(0001) using density functional theory
calculation. It is found that the intermediate hybridization between
BL and one QL $\beta-$In$_2$Se$_3$(0001) results in the formation of
bands with giant Rashba spin splitting in the large band gap of the
substrate. Furthermore the Rashba parameter $\alpha_R$ can be
increased significantly by tensile strain of substrate. Our findings
provide a good candidate substrate for BL growth to experimentally
realize spin splitting Rashba states with insignificant effect of
spin degenerate states from the substrate.
\end{abstract}

\maketitle

The last decade has witnessed intensive research efforts in
exploring the spin-orbital coupling driven surface/interface
electronic structures in heavy-atom containing system for
spintronics applications, such as Rashba-like spin splitting
\cite{Rashba1, Rashba2, Rashba3, Rashba4, Rashba5, Rashba6, Rashba7,
Rashba8, Rashba9}. Rashba-like spin splitting occurs as a result of
the combination of the electrostatic potential gradient at the
interface/surface and the large atomic spin-orbital coupling. From
the device point of view, for example spin field transistor
\cite{SFT}, the ability of spin manipulation depends on the strength
of the spin-orbital Rashba parameter and device room temperature
operation and miniaturization requires a large Rashba parameter.
Many attempts have been made in this direction. The earliest
experimental realization of the Rashba effect was in the asymmetric
quantum well formed in InGaAs/InAlAs heterostructure \cite{Rashba1}.
Nobel metal surfaces (e.g. Au \cite{Rashba2}, Ag \cite{Rashba3} and
Ir \cite{Rashba4} ) and sp-orbital heavy metal surfaces (e.g. Bi
\cite{Rashba5}, Sb \cite{Rashba6} and Pb \cite{Rashba7}) exhibited
larger Rashba splitting. For further enhancement of Rashba
splitting, Bi/Ag(111) surface alloy \cite{Rashba8} and
Bi-submonlayer on semiconductor surface with dangling bonds (e.g.
Si(111) \cite{Rashba9}) were experimentally shown to have Rashba
spliting in the giant regime. However, due to the large number of
spin degenerated electrons from the bulk metal and strong
hybridization between heavy adatoms and the dangling bonds in the
semiconductor substrate, it is the typical situation that the spin
split Rashba-like states are accompanied by large amount of spin
degenerated states, hindering the spintronics applications based on
the spin polarized Rashba states. Most recently the Te-terminated
surface of BiTeX (X=Cl, Br and I) \cite{BiTeX1, BiTeX2} was shown to
hold surface states with giant Rashba splitting completely in the
bulk band gap ranging from 0.4 eV to 0.8 eV.

Other than the Rashba spin-orbital physics in submonolayer
Bi+substrate system, the two monolayers of bilayer (BL) Bi(111) film
has been theoretically studied as a model system of two dimensional
topological insulator \cite{Bi1} in free-standing form with both
simplest crystal structure and large spin-orbital band gap
\cite{Bi2}. Unfortunately it is still challenging to grow ultrathin
Bi(111) on a semiconductor (e.g. Si(111)-(7x7)) substrate with
dangling bonds \cite{bismuth2, bismuth3, bismuth4, bismuth5,
bismuth6}. It was shown that Bi prefers to initially grow along the
[110] direction with black phosphorus buckling, and transforms to
the [111] direction after reaching a critical thickness of 7~BLs.
The strong interaction between the Bi thin-film and dangling bonds
from the substrate causes such an unwanted transformation. For this
reason, van der Waals-like (vdW) epitaxial growth \cite{vdW1, vdW2}
with reduced interfacial interaction may be used to grow ultrathin
Bi(111) to avoid such structural transition. Recently, Hirahara
\emph{el al.} \cite{bilayer1} and Miao \emph{et al.} \cite{bilayer2}
have confirmed this hypothesis and shown that one to several BLs Bi
(111) film can be grown on 3D Bi$_2$Se$_3$ and Bi$_2$Te$_3$
substrates. However, it was observed that the Bi(111) bilayer
Rashba-like electronic states are entangled with the topological
Dirac surface states from substrate Bi$_2$Se$_3$ and Bi$_2$Te$_3$ in
the same energy range. Another well known problem is that the
substrates Bi$_2$Se$_3$ and Bi$_2$Te$_3$ both have small
spin-orbital induced band gaps less than 0.35 eV. These two problems
prevent us from detecting the electronic states from more Bi layer
in substrate supported Bi(111) BL film.

In this work we propose to grow Bi (111) BL on
$\beta-$In$_2$Se$_3$(111) substrate, which has a large band gap of
1.2~eV. It has the same layered crystal structure as Bi$_2$Se$_3$
and Bi$_2$Te$_3$, that is, it consists of quintuple-layers (QL)
stacking in z direction. Within each QL In and Se are covalent
bonded and between QLs it is the vdW bonding. Our first-principles
calculations show that the Bi(111) BL can remain stable without
structural transformation on the $\beta-$In$_2$Se$_3$ substrate. The
interfacial interaction is in the desirable "intermediate" region,
slightly stronger than the vdW bond but substantially weaker than
chemical surface adsorption. Most importantly, a Rashba-like spin
splitting of Bi BL surface bands is created in the middle of
substrate band gap, which is induced by the interfacial charge
transfer. In particular, the "intermediate" interfacial interaction
plays an important role in inducing the formation of Rashba
splitting. Therefore, this system can reduce significantly the
effect of substrate carriers and has the three following advantages:
(1) the substrate has a large bulk band gap; (2) it has small
interaction with Bi to avoid strong hybridization between the
surface and bulk states; (3) it can induce a large spin-splitting in
Bi surface bands.
\begin{figure}
\epsfig{figure=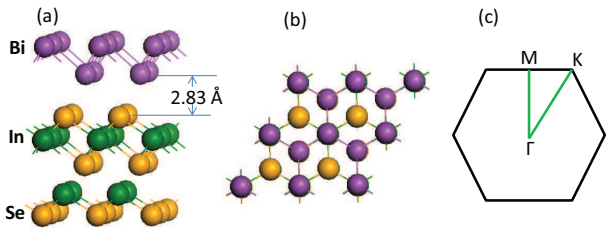,width=8cm}
\caption{(a) side view of the (2 $\times$ 2) supercell structure of
Bi(111) BL grown on one QL In$_2$Se$_3$ substrate. (b) top view of
the supercell. For clarity, ((2$\times$2)surface supercell is shown.
(c) 2D Brillouin zone of the hexagonal (1$\times$1) cell with high
symmetry directions $\Gamma$M and $\Gamma$K, where ~{\AA}
$|\Gamma$$M|=0.90${\AA} and $|\Gamma$$K|$=1.04{\AA}.  }

\end{figure}

Our calculations are carried out based on the density functional
theory with PAW pseudopotential \cite{paw} and PBE
exchange-correlation functional \cite{pbe} using VASP package
\cite{vasp}. The bulk $\beta$-In$_2$Se$_3$ in-plane lattice constant
is $4.02{\AA}$ from our calculation, which is very close to the
experimental value \cite{sIn2Se3}. The substrate is simulated by one
QL (1$\times$1))$\beta-$In$_2$Se$_3$(1111) with the bulk in-plane
lattice constant and additional vacuum of more than 20{\AA} normal
to the surface as shown in Fig. 1(a) of side view and Fig. 1(b) of
top view. During the structural relaxation, all the atoms are
allowed to relax until the forces are smaller than 0.01 eV/{\AA}.
600 eV kinetic energy cutoff and 11$\times$11$\times$1 Gamma
centered k-mesh sampling are adopted for total energy convergence.
After structural relaxation the vertical distance between top Bi
bilayer and bottom substrate is $\sim$~2.83~$\AA$, and the
inter-layer binding energy is $\sim$ 0.3~eV/surface-unit-cell. This
indicates that the interfacial interaction is much stronger than vdW
bond but significantly weaker than typical chemical bond.

\begin{figure}
\epsfig{figure=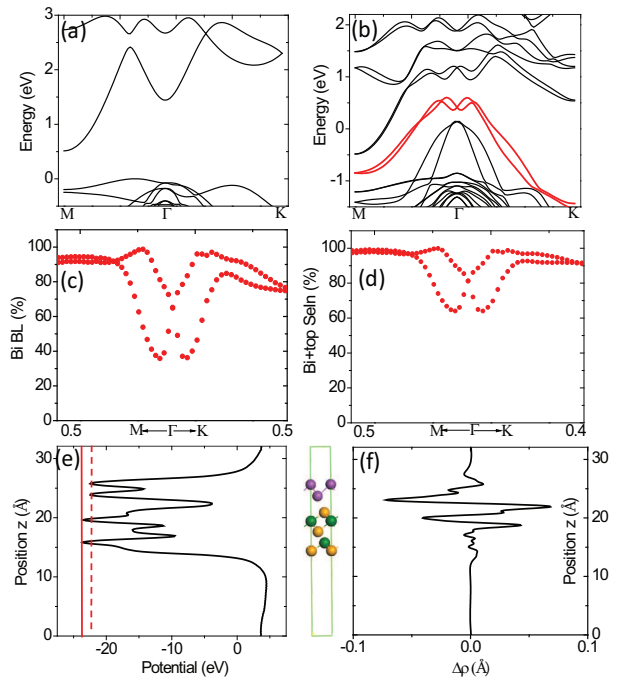,width=8cm} \caption{(a) Band structure of
bare one-QL In$_2$Se$_3$ along $M-\Gamma-K$. (b) Band structure of
Bi(111) bilayer on one QL In$_2$Se$_3$ along $M-\Gamma-K$. The
red-color marks the bands we are interested in with Large
Rashba-like spin splitting. (c) and (d) The charge localization of
the red-color marked two bands in (a) as a function of k-points in
the vicinity of $\Gamma$ along the $M-\Gamma-K$ from top Bi(111) BL
and in top Bi(111) BL+the first Se+In layers of the substrate one QL
In$_2$Se$_3$, respectively. (e) Inplane averaged Kohn-Sham potential
as a function of position z. The sold red vertical line and the
dashed vertical line show the potential valley of Bi(111) BL and one
QL In$_2$Se$_3$, respectively. (f) Charge transfer between Bi(111)
BL and one QL In$_2$Se$_3$ along the z direction. It shows charge
transfer at the interface from bottom Bi atomic layer to the top of
one QL In$_2$Se$_3$. In-between Fig. (e) and (f) is the supercell of
Bi(111) bilayer+one QL In$_2$Se$_3$ in order to show the
corresponding position in the supercell for a given z value.}
\end{figure}

\begin{figure*}
\epsfig{figure=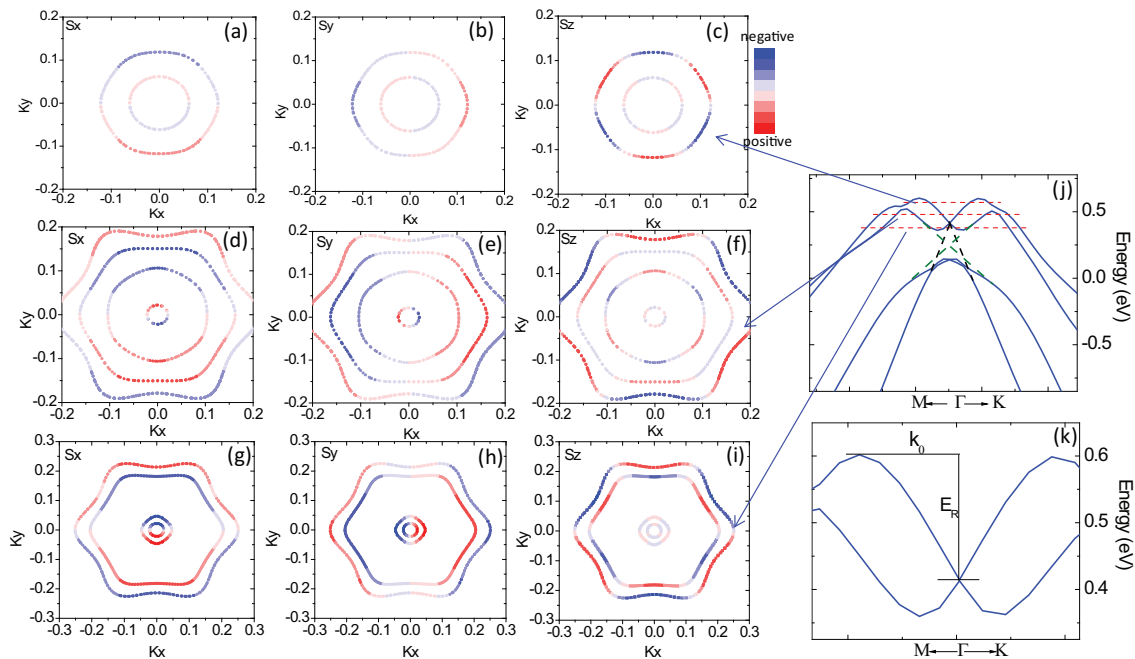,width=15cm} \caption{(a)-(c) The spin
texture of the x, y, z components for isoenergy surface 150 meV
above, (d)-(f) 60 meV above and (g)-(i) 40 meV below the spin
degenerate point at $\Gamma$. (j) The zoomed-in view of the bands
with Rashba-like spin splitting in the gap of one-QL In$_2$Se$_3$.
The dashed red lines point out the position of isoenergy surface 150
meV above, 60 meV above and 40 meV below the spin degenerate point
at $\Gamma$, respectively. The green and the red dashed lines
indicate the band dispersion trend if the interaction at the avoid
band crossing point is turned off. (k) Labeling of offset k vector
(k$_0$) and Rashba energy (E$_R$).}
\end{figure*}

Fig. 2(a) and 2(b) show the band structures of bare one QL
In$_2$Se$_3$ and Bi BL on In$_2$Se$_3$ along $M-\Gamma-K$
directions, respectively. In Fig. 2(a) we found that bare one QL
In$_2$Se$_3$ has an indirect band gap of 0.5 eV with the conduction
band minimum (CBM) at M point and that the band gap at other
k-points away from M is larger than 1.0 eV. Most interestingly
because the QL surface is from the cleavage through the vdW gap and
does not break chemical bonds, there are no metallic surface states
formed at the surface and the QL substrate still remains a
semiconductor. We have verified that this behavior is not due to
quantum confinement effect with our complementary calculation of the
band structure of In$_2$Se$_3$(0001) substrate with significantly
larger number of QLs, which shows that it is still a semiconductor
except slight different band gap from that of one QL. In Fig. 2(b)
the bands marked with red color shows Rashba-like splitting in the
vicinity of $\Gamma$ point. Due to the spatial inversion symmetry,
all bands in free-standing Bi BL are two-fold degenerated. If the
interaction between Bi BL and the substrate is negligibly weak, the
Bi BL will be chemically decoupled with the substrate and spatial
inversion symmetry of Bi Bi is locally retained, and then Bi bands
should not be split. Therefore, the inter-layer interaction must
play a key role for the Bi band-splitting. It is well know that
LDA/GGA underestimates the band gap of semiconductor. This may
affect the band gap of the one-QL In$_2$Se$_3$ and also the energy
position of the Rashba-like bands in the gap region. We thus used
hybrid functional \cite{HSE} to do the similar band structure
calculations for bare one QL In$_2$Se$_3$ and Bi BL on In$_2$Se$_3$.
We found that the energy gap becomes around 1.2 eV due to the
valence band shift to the lower energy, but the relative position
between the Rashba-like bands and the substrate conduction bands
does not change. Therefore in the following all the results are from
standard DFT calculations with PBE functional.

To get a physical insight of such interfacial interaction, we
calculate the charge localization of the two Rashba split bands as a
function of k-points along the $M-\Gamma-K$ as shown in Figs. 2(c)
and (d) for charge distribution only from top Bi BL and top Bi BL +
the top Se/In atomic layers, respectively. Around $\Gamma$ point,
the states have comparable distributions in both Bi BL and
substrate. This indicates that a sizable hybridization between Bi BL
and substrate may occur for the states around $\Gamma$. Away from
$\Gamma$, the charge almost all localizes in Bi BL, having little in
the substrate. In Fig. 2(e) we plot the in-plane averaged electron
potential along z direction. It is seen that the potential-well at
Bi BL is asymmetric with the top potential higher at the interface
between top Bi layer and vacuum than the bottom potential at the
interface between bottom Bi layer and the substrate. This causes the
top Bi layer is not equivalent to the bottom Bi layer and thus
breaks the inversion symmetry that holds in free-standing Bi BL. The
solid and dashed lines indicate the potential minima at the
substrate QL and Bi BL respectively with substrate QL potential
minimum lower than Bi BL. This indicates that the charge transfer
between Bi-BL and the substrate will be from Bi-BL to the substrate.
We then calculate the charge transfer ($\Delta\rho=\int
dxdy(\rho_{Bi BL+In_2Se_3}(x,y,z)-\rho_{In_2Se_3}(x,y,z)-\rho_{Bi
BL}(x,y,z)$) between Bi BL and the substrate along z direction in
Fig. 2(f). Clearly, there is a substantial charge transfer at the
interface. The Bi BL acts as an electron-donor and the In$_2$Se$_3$
substrate as an electron-acceptor, with the electrons transferring
from the former to the latter. This is consistent with the result
from the analysis of electron potential in Fig. 2(e). This charge
transfer generates a large internal electric field at the interface
region. The field direction points from the Bi to In$_2$Se$_3$ and
field strength is estimated as high as $\sim$0.40 $V/{\AA}$.

Such charge transfer induced band splitting is very similar to the
external electric field induces Rashba band splitting in the surface
state of topological insulators \cite{chenli, SbTe}. To clearly see
this similarity, in Fig. 3(j) and (k) we zoom-in the Rashba-split
bands from Fig. 2(b) around $\Gamma$ point. The resulting band
structure is very different from that of single pair of
free-electron Rashba-split bands and The gap openings around the
avoid-crossing k-points are found. We use dashed red and blue lines
to track back possible band structure for the situation without
interaction between the bands. This gives two pairs of downward
Rashba bands. Because of time reversal symmetry the the splitting
bands are degenerated at $\Gamma$ point. The estimated Rashba energy
($E_R$), offset k vector (k$_0$) from $\Gamma$ and Rashba parameter
($\alpha_R=2E_R/k_0$) are 182 meV, 0.10 $\cdot{\AA}^{-1}$ and 3.66
$eV\cdot{\AA}$, respectively. Here, we would like to emphasize three
characters of the Rashba-like splitting in our system. First the
$E_R$ is comparable to the largest value that has ever been observed
in Bi/Ag(111) \cite{Rashba8}. Second in Fig. 3(a)-(i) we show the
spin textures for x, y and z component of spin polarization vector
for the three isoenergy surfaces marked by red dashed lines as in in
Fig. 3(j) (150 meV, 60 meV above and 40meV below the degenerate
point at $\Gamma$, respectively). The spin polarization is  defined
as \cite{polar}:
$\overrightarrow{p}(\overrightarrow{k})=[<S_x(\overrightarrow{k})>,
<S_y(\overrightarrow{k})>, <S_z(\overrightarrow{k})>]$, where
$<S_\alpha(\overrightarrow{k})>=<\psi(\overrightarrow{k})|\sigma_\alpha|\psi(\overrightarrow{k})>,
(\alpha=x,y,z)$. Because there are four bands, it displays four
isoenergy surfaces in Fig. (d)-(i). The isosurface is only isotropic
having a circular shape for small k vector (less than 0.15~
{\AA}$^{-1}$) around $\Gamma$ and begins to gain some anisotropy for
larger k vector having hexagonal shape. For the Sz component it has
very small value along the path of small k-vector but it acquires
sizeable value with increasing k-vector. This is different from the
isoenergy surfaces and spin textures of free-electron Rashba model
with only potential gradient along z direction, where the isoenergy
surface is isotroic and the Sz component is zero with only in-plane
spin components. The anisotropic isoenergy surface and out-of-plane
spin component can be attributed to the in-plane potential gradient
imposed by the periodic crystal field. Similar behavior has been
observed in many studies for Bi/Ag(111) surface alloying system
\cite{Rashba8} and 3D topological Dirac surface state in
Bi$_2$Te$_3$ \cite{SzTI}. Third the Sz component oscillates
periodically around the isosurface with a period of $120^\circ$
which reflects the in-plane threefold rotational symmetry and have
the opposite phase above and below the spin degenerate point at
$\Gamma$, similar to the topological surface states of 3D
Bi$_2$Se$_3$ \cite{osig}.

\begin{figure}
\epsfig{figure=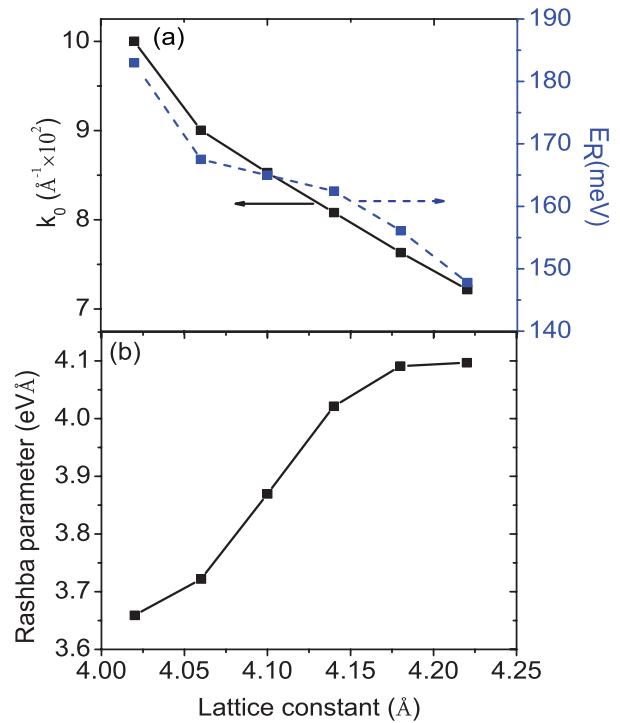,width=8cm} \caption{(a) Lattice constant
dependence of k$_0$ and E$_R$. (b) Lattice constant dependence
Rashba parameter $\alpha_R$.}
\end{figure}

Lastly we notice that the In$_2$Se$_3$ substrate used is only one QL
thick, hence it may be readily strained. Under substrate tensile
strain, Bi BL can have less lattice mismatch with In$_2$Se$_3$ and
in turn the Rashba parameter may be tuned with strain. Fig. 4(a)
shows the strain dependence of k$_0$ and E$_R$ with a tensile strain
from $0\%$ to $5\%$ corresponding to inplane lattice constant from
4.02 $\cdot{\AA}$ to 4.22 $\cdot{\AA}$. They both decrease with
increasing lattice constant with k$_0$ decreased by $28\%$ and E$_R$
decreased by $20\%$. Fig. 4(b) shows that $\alpha_R$ increases with
increasing lattice constant and reaches 4.22 $eV\cdot{\AA}$ at
inplane lattice constant of 4.22 $\cdot{\AA}$. Such a trend of
increase may be related to shorter vertical separation between Bi BL
and the substrate which deceases from 2.83 $\cdot{\AA}$ at 4.02
$\cdot{\AA}$ lattice constant to 2.51 $\cdot{\AA}$ at 4.02
$\cdot{\AA}$ lattice constant, so that Bi BL feels larger electric
field which induces larger $\alpha_R$.

The Indium Selenides family also has a few allotropes with different
structures such as hexagonal and rhombohedral structures, and
compositions such as InSe and In$_2$Se$_3$  \cite{vdW1,InSe1, InSe2,
InSe3}. All of these allotropes have the similar layered structure
with vdW type inter-layer bonding, which may give rise to the
"intermediate" interaction between the Bi BL and substrate. In
addition, the band gaps of these allotropes range from 1.2 eV to 1.8
eV. Thus, we speculate that Bi BL may also grow on these substrates
to create the Rashba spin splitting. The experimental work is
expected to verify our proposal.

To summarize, we have investigated the electronic structures of
Bi(111) BL grown on substrate of one-QL In$_2$Se$_3$, which has a
large band gap of 1.2 eV and vdW like surface cleavage necessary for
the stablization of Bi in (111) BL phase. Overlying Bi BL is
predicted to have giant Rashba-like spin splitting in the substrate
band gap. The interlayer interaction between Bi BL and In$_2$Se$_3$
is of "intermediate" strength, which plays two important roles: (1)
strong enough to induce sizable charge transfer and hence the
Rashba-like strong spin splitting; (2) weak enough to avoid strong
hybridization between Bi overlayer and substrate states and separate
the substrate states from being in the energy range. These findings
in this work may provide a good candidate substrate for Bi(111) film
to realize spin splitting Rashba states with insignificant effect of
spin degenerate states from the substrate.

This work was supported by NSF MRSEC (Grant No. DMR-1121252) and
DOE-BES (Grant No. DE-FG02-04ER46148). We thank the CHPC at
University of Utah and NERSC for providing the computing resources.



\begin{references}
\bibitem{Rashba1}
JunSaku Nitta, Tatsushi Akazaki, and Hideaki Takayanagi, Phys. Rev.
Lett. {\bf 78}, 1335 (1997).

\bibitem{Rashba2}
S. LaShell, B. A. McDougall, and E. Jensen, Phys. Rev. Lett. {\bf
77}, 3419 (1996).

\bibitem{Rashba3}
D. Popovic, F. Reinert, S. Hufner, V. G. Grigoryan, M. Springborg,
H. Cercellier, Y. Fagot-Revurat, B. Kierren and D. Malterre, Phys.
Rev. B {\bf 72}, 045419 (2005).

\bibitem{Rashba4}
A. Varykhalov, D. Marchenko, M. R. Scholz, E. D. L. Rienks, T. K.
Kim, G. Bihlmayer, J. Sanchez-Barriga, and O. Rader, Phys. Rev.
Lett. {\bf 108}, 066804 (2012).

\bibitem{Rashba5}
Yu. M. Koroteev, G. Bihlmayer, J. E. Gayone, E. V. Chulkov, S.
Blugel, P. M. Echenique, and Ph. Hofmann, Phys. Rev. Lett. {\bf 93},
046403 (2004).

\bibitem{Rashba6}
K. Sugawara, T. Sato, S. Souma, T. Takahashi, M. Arai, and T.
Sasaki, Phys. Rev. Lett. {\bf 96}, 046411 (2006).

\bibitem{Rashba7}
J. Hugo Dil, Fabian Meier, Jorge Lobo-Checa, Luc Patthey, Gustav
Bihlmayer, and Jurg Osterwalder, Phys. Rev. Lett. {\bf 101}, 266802
(2008).

\bibitem{Rashba8}
Christian R. Ast, Jurgen Henk, Arthur Ernst, Luca Moreschini,
Mihaela C. Falub, Daniela Pacile, Patrick Bruno, Klaus Kern, and
Macro Grioni, Phys. Rev. Lett. {\bf 98}, 186807 (2007).

\bibitem{Rashba9}
I. Gierz, T. Suzuki, E. Frantzeskakis, S. Pons, S. Ostanin, A.
Ernst, J. Henk, M. Grioni, K. Kern, and C. R. Ast, Phys. Rev. Lett.
{\bf 103}, 046803 (2009).

Haijun Zhang, Chao-Xing Liu, Xiao-Liang Qi, Xi Dai, Zhong Fang, and
Shou-Cheng Zhang, Nat. Phys. {\bf 5}, 438 (2009).

B. Andrei Bernevig, Taylor L. Hughes, and Shou-Cheng Zhang, Science
{\bf 314}, 1757 (2006).

\bibitem{SFT}
S. Datta, and Das, Appl. Phys. Lett. {\bf 56}, 665 (1990).

\bibitem{BiTeX1}
S. V. Eremeev, I. A. Nechaev, Yu. M. Koroteev, P. M. Echenique, and
E. V. Chulkov, Phys. Rev. Lett. {\bf 108}, 246802 (2012).

\bibitem{BiTeX2}.
M. Sakano, M. S. Bahramy, A. Katayama, T. Shimojima, H. Murakawa, Y.
Kaneko, W. Malaeb, S. Shin, K. Ono, H. Kumigashira, R. Arita, N.
Nagaosa, H. Y. Hwang, Y. Tokura, and K. Ishizaka, Phys. Rev. Lett.
{\bf 110}, 107204 (2013).


\bibitem{Bi1}
S. Murakami, Phys. Rev. Lett. {\bf 97}, 236805 (2006).
\bibitem{Bi2}
Zheng Liu, Chao-Xing Liu, Yong-Shi Wu, Wen-Hui Duan, Feng Liu, and
Jian Wu, Phys. Rev. Lett. {\bf 107}, 136805 (2011).

\bibitem{bismuth2}
T. Nagao, \emph{et al.}, Phys. Rev. Lett. {\bf 93}, 105501 (2004).

\bibitem{bismuth3}
 S.Yaginuma, \emph{et al.}, Surf. Sci. {\bf 601}, 3593 (2007).

\bibitem{bismuth4}
J. T. Sadowski, \emph{et al.}, J. Appl. Phys. {\bf 99}, 014904
(2006).

\bibitem{bismuth5}
T. Nagao, \emph{et al.}, Surf. Sci. {\bf 590}, L247 (2005).

\bibitem{bismuth6}
S. A. Scott, \emph{et al.}, Surf. Sci. {\bf 587}, 175 (2005).

\bibitem{vdW1}
A. Klein, O. Lang, R. Schlaf, C. Pettenkofer and W. Jaegermann,
Phys. Rev. Lett. {\bf 80}, 361 (1998).

\bibitem{vdW2} Atsushi Koma, J. Cryst. Growth {\bf 201}, 236 (1999).

\bibitem{bilayer1}
Toru Hirahara, Gustav Bihlmayer, Yusuke Sakamoto, Manabu Yamada,
Hidetoshi Miyazaki, Shin-ichi Kimura, Stefan Blugel and Shuji
Hasegawa, Phys. Rev. Lett. {\bf 107}, 166801 (2001).

\bibitem{bilayer2}
Lin Miao, Z. F. Wang, Wenmei Ming, Meng-Yu Yao, Meixiao Wang, Fang
Yang, Y. R. Song, Fengfeng Zhu, Alexei V. Fedorov, Z. Sun, C. L.
Gao, Canhua Liu, Qi-Kun Xue, Chao-Xing Liu, Feng Liu, Dong Qian, and
Jin-Feng Jia, PNAS {\bf 110}, 2758 (2013).



\bibitem{paw}
G. Kresse and D. Joubert, Phys. Rev. B {\bf 59}, 1758 (1999).

\bibitem{pbe}
J. P. Perdew, K. Burke, and M. Ernzerhof, Phys. Rev. Lett. {\bf 77},
3865 (1996).

\bibitem{vasp}
G. Kresse and J. Furthm$\ddot{u}$ler, Comput. Mat. Sci. {\bf 6},
15(1996).

\bibitem{sIn2Se3}
Kozo Osamura, and Yataro Murakami, J. Phys. Soc. Japan {\bf 21},
1848 (1966).

\bibitem{HSE}
J. Heyd, G. E. Scuseria, and M. Ernzerhof, J. Chem. Phys. {\bf 118},
8207 (2003); {\bf 124}, 219906 (2006).

\bibitem{SbTe}
Minsung Kim, Choog H. Kim, Heung-Sik Kim and Jisoon Ihm, PNAS {\bf
109}, 671 (2011).

\bibitem{chenli}
Li Chen, Z. F. Wang, and Feng Liu, Phys. Rev. B {\bf 87}, 235420
(2013).

\bibitem{polar}
O. Yazyev, \emph{et al.}, Phys. Rev. Lett. {\bf 105}, 266806 (2010).

\bibitem{SzTI}
S, Souma, \emph{et al.}, Phys. Rev. Lett. {\bf 106}, 216803 (2011).

\bibitem{osig}
Y. Zhao, \emph{et al.}, Nano Lett. {\bf 11}, 2088 (2011)







\bibitem{InSe1}
P. Gomes da Costa, \emph{et al.}, Phys. Rev. B {\bf 48}, 14135
(1993)

\bibitem{InSe2}
Jiping Ye, Sigeo Soeda, Yoshio Narkamura and Osamu Nittono, Jpn. J.
Appl. Phys. {\bf 37}, 4264 (1998);

\bibitem{InSe3}
Hailin Peng, Chong Xie, David T. Schoen and Yi Cui, Nano lett. {\bf
8}, 1511 (2008)


\end{references}
\end{document}